\documentstyle[epsfig]{elsart}

\begin{document}
\begin{frontmatter}

%\sloppy
%\thispagestyle{empty}

%%\setcounter{page}{0}

%\mbox{}
%\vspace*{\fill}
%\begin{center}
%{\LARGE\bf Comments on multiple scattering of high-energy muons in thick 
%layers} \\

\title{Comments on multiple scattering of high-energy muons in thick 
layers}

%\vspace{2mm}
%{\LARGE\bf title continued e.g.}\\

%\vspace{2em}
%\large
\author[INR]{A.V.~Butkevich\corauthref{cor}}, 
\corauth[cor]{ Corresponding author. Tel.: 07-095-334-0188; 
Fax: +07-095-334-0184.}
\ead{butkevic@al20.inr.troitsk.ru}
\author[MIFI]{R.P.~Kokoulin},
\author[INR]{G.V.~Matushko} and 
\author[INR]{S.P.~Mikheyev}
\address[INR]{Institute for Nuclear Research of Russian Academy of Science,
60th October Anniversary prospect, 7a, Moscow 117312, Russia.}
\address[MIFI]{Moscow State Engineering Physics Institute 
(Technical University), Kashirskoe Shosse, 31, Moscow 115409, Russia.}

%%\\
%%\vspace{2em}
%%\small
%{\it $^1$ Institute for Nuclear Research of Russian Academy of Science,}
%{\it 60th October Anniversary prospect,7a, Moscow 117312,
% Russia}\\

%\vspace{2em}
%{\it $^2$ Moscow State Engineering Physics Institute (Technical University),}
%{\it Kashirskoe Shosse, 31, Moscow 115409, Russia}\\

%\vspace{2em}

%\footnotetext{$^*$ Corresponding author. Tel.: 07-095-334-0188; 
%Fax: +07-095-334-0184.\\
%{\it E-mail address}: butkevic@al20.inr.troitsk.ru}
%\end{center}
%\vspace*{\fill}
%

\begin{abstract}
%\section*{Abstract}
%\noindent
We describe two independent methods to calculate the angular
 distribution of muons after traversing a thick scatterer due to multiple
Coulomb scattering. Both methods take into account the nuclear size effect. 
We demonstrate a necessity to account for the nucleus extension as well as 
incoherent scattering on atomic electrons to describe the muon scattering 
at large angles in thick matter layers. The results of the two methods of 
calculations are in good agreement.\\
\


\end{abstract}
\end{frontmatter}
\vspace*{\fill}
\newpage
%
%%%%%%%%%%%%%%%%%%%%%%%%%%%%%%%%%%%%%%%%%%%%%%%%%%%%%%%%%%%%%%%%%%%%%%%%%%
\section{Introduction}
\label{sect1}
%_______________________________________________________
%Now your test starts.
\hspace*{0.6cm}
Multiple scattering of muons in the field of atoms is of interest for numerous
applications related to muon transport in matter, in particular, for 
simulation of the response of high-energy particle detectors (including 
stochastic particle deviation in magnetized steel spectrometers), radiation 
protection tasks at accelerators, muon-induced background evaluations for 
cosmic ray neutrino experiments, etc. In many cases, reliable estimations 
of angular and lateral muon distribution functions at the level of probability
 of about $10^{-2} - 10^{-4}$ and lower are necessary. Furthermore, renewed 
interest to particle momentum evaluation on the basis of precise measurements
of multiple scattering effect (utilising the scales and coordinate accuracies 
of novel muon and neutrino detectors) calls for adequate accuracies in the 
theoretical description of the phenomenon.\\
\hspace*{0.6cm}
Several multiple scattering theories have been published (see review [1]) 
which are concerned with angular distribution of particles passing without 
substantial loss of energy $\Delta E \ll E$ using the small-angle approximation
of scattering angle $\chi$. This approximation assumes that $\chi$ is 
small and consists in replacing $\sin(\chi) \approx \tan(\chi) \approx 
\chi$ and the upper limit $\pi$ for $\chi$ by infinity. 
Gaussian distribution function is widely used for various estimations, however
it can hardly serve as a satisfactory tool for consideration of low 
probability region (large deflection angles), especially at moderate absorber
thickness. On the other hand, Moliere theory [2],[3] of multiple 
scattering, the results of which are employed nowadays in most of the 
transport codes, cannot be used for description of high-energy muon scattering
either. The reason is that, due to its lagre mass, muon is capable of passing 
large thicknesses of material, and hence the probability of scatters with 
impact parameters comparable with nuclear size is not small. Moliere theory 
(as well as other approaches developed within the point-like  nucleus 
approximation) heavily overestimates the probability of large angle 
deflections.\\
\hspace*{0.6cm}
Several calculations have been carried out to take into account the charge 
distribution in nucleus.  Two methods [4] for evaluating 
multiple scattering on extended nuclei were given, one of which is of 
interest as a general method for any charge distribution. The expression 
for angular distribution was obtained in Ref.[5] under the assumption that the 
charge distribution in nuclei and proton is Gaussian. Unfortunately, there 
are several misprints in the work [5].\\ 
\hspace*{0.6cm}
Moliere theory regarded elastic collisions against the screened Coulomb 
field of atomic nuclei only. Collisions with Z atomic electrons also 
contribute to multiple scattering, especially in light elements. This effect
is often introduced by replacing $Z^2$ with $Z(Z+1)$, which serves only 
to estimate the order of magnitude of the effect of inelastic scattering. 
The account of inelastic collisions in the Moliere theory has been 
done in Ref.[6] for electrons and heavy particles.\\ 
\hspace*{0.6cm}
In this paper, we present two methods for calculation of the muon angular
distribution function due to multiple Coulomb scattering (MCS) in matter.
The procedures take into account the nuclear size effect and contribution
of incoherent scattering on atomic electrons.\\
\hspace*{0.6cm}
The paper is organized as follows. In Sec.2, we give the general relations and 
definitions used for description of multiple scattering of muons in
the field of atoms. A mixed Monte Carlo approach to the MCS consideration 
is described in Sec.3. In Sec.4, contribution
 of incoherent scattering on atomic electrons and the multiple scattering 
in mixtures and compounds are discussed. Basic expressions of Moliere 
multiple scattering theory are given in Sec.5, whereas the equations of 
modified Moliere theory taking into account nuclear size effect are 
described in Sec.6. At the end of this section, we compare the angular 
distributions of muons obtained by Monte Carlo method and analytically. In the
conclusion, the main results of the paper are summarized.
%____________________________________________________ 
\section{General relations and definitions}
\label{sect2}
\hspace*{0.6cm}
Differential cross section (per gram of matter) for scattering in Coulomb 
field of nuclei in the small angle approximation can be written as follows
[7]:
%%%%%%%%%%%%%%%%%%%%%%%%%%%%%%%%%%%%%%%%%%%%%%%
\begin{equation}
{\d}\sigma_0(\chi)=8\pi N_{\mathrm{Av}}\frac{Z^2}{A}\left (\frac{e^2}{pv}
\right )^2\frac{{\d}\chi}{\chi^3} =8\pi N_{\mathrm{Av}}\frac{Z^2}{A}\left 
(\frac{r_em_e}{p\beta}\right  )^2\frac{{\d}\chi}{\chi^3},                %1 
\end{equation}
%%%%%%%%%%%%%%%%%%%%%%%%%%%%%%%%%%%%%%%%%%%%%%
where $N_{\mathrm{Av}}$ is the Avogadro number, $A$ is the atomic weight, 
$p$ and $v$ 
are momentum and velocity of the particle, respectively. In the second 
relation, classical electron radius $r_e$ and electron mass $m_e$ are 
utilised as natural units of distance and energy. Hereafter, whenever it 
does not cause misunderstanding, we assume $\hbar=c=1$. For high-energy 
($E \sim p$, $\beta \sim 1$) muon scattering on heavy target through small 
angles, the transverse momentum transferred to the particle $q=E\chi$, and 
practically coincides with the total 3-dimensional momentum transfer. The 
cross section may be re-written as:
%%%%%%%%%%%%%%%%%%%%%%%%%%%%%%%%%%%%%%%%%%%%%%%
\begin{equation}
{\d}\sigma_0(q)=8\pi N_{\mathrm{Av}}\frac{Z^2}{A}\left (r_em_e\right )^2
\frac{{\d} q}{q^3}. %2
\end{equation}
%%%%%%%%%%%%%%%%%%%%%%%%%%%%%%%%%%%%%%%%%%%%%%
Noteworthy, in this ultrarelativistic regime the cross section depends only 
on transverse momentum and not on the particle energy. The integral cross 
section for scattering with transverse momenta greater than $q$ is:
%%%%%%%%%%%%%%%%%%%%%%%%%%%%%%%%%%%%%%%%%%%%%%%
\begin{equation}
\sigma_0^{int}(q)=4\pi N_{\mathrm{Av}}\frac{Z^2}{A}\left (\frac{r_em_e}{q}
\right)^2.  %3
\end{equation}
%%%%%%%%%%%%%%%%%%%%%%%%%%%%%%%%%%%%%%%%%%%%%%
Differential cross section for extended nuclei (taking into account also 
atomic screening) in the Born approximation may be described by means of 
formfactors:
%%%%%%%%%%%%%%%%%%%%%%%%%%%%%%%%%%%%%%%%%%%%%%
\begin{equation}
{\d}\sigma(q)={\d}\sigma_0(q)\left \{\left (F_N-F_a\right )^2+\frac{1}{Z}\left 
(1-F_N^2\right )F_p^2\right \},     %4
\end{equation}
%%%%%%%%%%%%%%%%%%%%%%%%%%%%%%%%%%%%%%%%%%%%%%
where $F_a(q)$, $F_N(q)$, and $F_p(q)$ are elastic atomic, nuclear, and 
proton formfactors, respectively. The second term in braces represents 
the (order-of-magnitude) correction for quasi-elastic processes (with 
excitation or disintegration of the nucleus), if incoherent scattering 
of the projectile on individual protons inside nucleus occurs. For hydrogen 
($Z=$1) this term should be omitted.\\
\hspace*{0.6cm}       
Formfactors are normalised in such a way that $F(0)=1$ and $F(\infty)=0$. 
A big difference between the sizes of atom and nucleus allows, as a rule, 
to consider the influence of atomic and nuclear formfactor separately: in 
the region where $F_a$ is essential, $F_N=1$; on the contrary, when $F_N$ 
appreciably deviates from the unit, $F_a$ is negligibly small. Hence, in 
a wide range of momentum transfers (two - three orders of magnitude, 
depending on the substance), and, as a consequence, over 4 - 5 decades in 
interaction frequency determined by Eq.(3), individual scatters obey a 
simple Rutherford law.\\
\hspace*{0.6cm}
The functional dependence of the formfactors is determined by the charge 
density distribution inside the object. For light and medium nuclei, a 
sufficiently good description may be reached on the basis of two-parameter 
distribution function suggested by Fermi. However, the use of accurate charge 
distributions entails serious computational difficulties and cannot be done 
in analytical form. Therefore, reasonable approximations are often used 
(for example, Gaussian or exponential distribution of charge density). 
For consideration of multiple scattering process, such approximations
are justified by the fact that at moderate transferred momenta  
the same expansion can be written for all formfactors
%%%%%%%%%%%%%%%%%%%%%%%%%%%%%%%%%%%%%%%%%%%%%%
\begin{equation}
F_N(q)\simeq 1-\left (qR_N\right )^2/6,                                %5
\end{equation}
%%%%%%%%%%%%%%%%%%%%%%%%%%%%%%%%%%%%%%%%%%%%%%
where $R_N$ is rms radius of charge distribution. An appreciable 
difference between different formfactors appears only in rare collisions with 
large $q$. Analytically, within the framework of Born approximation [8]
%%%%%%%%%%%%%%%%%%%%%%%%%%%%%%%%%%%%%%%%%%%%%%
\begin{equation}
F_N(q)=\exp\left [-\frac{(qR_N)^2}{6}\right ]~~~ \mbox{and}~~~ 
F_N(q)=\left [1+\frac{(qR_N)^2}{12}\right ]^{-2}          %6
\end{equation}
%%%%%%%%%%%%%%%%%%%%%%%%%%%%%%%%%%%%%%%%%%%%%%
for Gaussian and exponential charge distributions, respectively. Rms-radii 
for practically all elements (and, moreover, for numerous isotopes) can be 
found elsewhere (e.g., see the compillation in Ref.[9]). For light and 
medium nuclei, the $A$-dependence of $R_N$ may be parameterized as
%%%%%%%%%%%%%%%%%%%%%%%%%%%%%%%%%%%%%%%%%%%%%%
\begin{equation}
R_N=1.27A^{0.27} {\mathrm{~fm}}~~~ \mbox{or}~~~ q_N=1/R_N=155A^{-0.27} 
{\mathrm{~MeV}}.                                           %(7)
\end{equation} 
%%%%%%%%%%%%%%%%%%%%%%%%%%%%%%%%%%%%%%%%%%%%%
For proton, $R_p=0.85$ fm, and $q_p=232$ MeV.\\
\hspace*{0.6cm} 
The important parameter in consideration of multiple scattering phenomenon is 
a so-called characteristic value $q_c$ of transverse momentum, which 
determines on the average one scatter with $q>q_c$ at a given layer thickness
X(g/cm$^2$):
%%%%%%%%%%%%%%%%%%%%%%%%%%%%%%%%%%%%%%%%%%%%%%%
\begin{equation}
q_c^2=4\pi N_{\mathrm{Av}}\frac{Z^2}{A}\left (r_em_e\right )^2X.            %8
\end{equation}
%%%%%%%%%%%%%%%%%%%%%%%%%%%%%%%%%%%%%%%%%%%%%%
Sometimes, the following regions are distinguished in the 
scattering process: multiple scattering ($q\ll q_c$, numerous scatters), 
plural scattering (several, but few, scatters with $q\sim q_c$), and single 
scattering (low probability scatters with $q\gg q_c$); in the latter domain, 
scattering angle distribution function approaches the behaviour of the 
differential cross section.\\
\hspace*{0.6cm}
Statistical consideration of multiple scattering is appropriate when 
$q_c\gg 1/R_a$, where $R_a$ is the effective radius of the atomic electron 
charge distribution:
%%%%%%%%%%%%%%%%%%%%%%%%%%%%%%%%%%%%%%%%%%%%%%%
\begin{equation}
R_a=1/q_a=\frac{183Z^{-1/3}}{2.718m_e}.            %9
\end{equation}
%%%%%%%%%%%%%%%%%%%%%%%%%%%%%%%%%%%%%%%%%%%%%%
Hence, the applicability of theory is usually limited to thickness of
layers:
%%%%%%%%%%%%%%%%%%%%%%%%%%%%%%%%%%%%%%%%%%%%%%%
\begin{equation}
X\gg 4\cdot 10^{-4}A/Z^{4/3}  {\mathrm{~g/cm^2}}.            %10
\end{equation}
%%%%%%%%%%%%%%%%%%%%%%%%%%%%%%%%%%%%%%%%%%%%%%
%_____________________________________________
\section{Monte Carlo approach}
\label{sect3}
\hspace*{0.6cm}
The idea of MCS consideration by means of a 
mixed Monte Carlo technique (see, e.g., Ref.[10]) consists in the 
separation of multiple scattering through small angles, or with small 
transferred momenta (which are treated in Gaussian approximation), and 
of scatters in the region of large $q$, which are simulated explicitely. 
This approach can be conveniently used for calculations with accurate cross 
section in high $q$ region (in particular, including nuclear formfactors). 
The boundary $q_b$ between these two domains has to satisfy the following 
inequalities:
%%%%%%%%%%%%%%%%%%%%%%%%%%%%%%%%%%%%%%%%%%%%%%%
\begin{equation}
q_a\ll q_b \ll q_N; ~~~~ q_b\le q_c.               %11
\end{equation}
%%%%%%%%%%%%%%%%%%%%%%%%%%%%%%%%%%%%%%%%%%%%%%
The first of these two conditions ensures Rutherford behaviour of the 
cross section around $q_b$, while the second one provides the regime of 
plural or multiple interactions for scatters that are simulated individually.\\
\hspace*{0.6cm}
In order to calculate the rms-deviation for small angle scatters with 
 $q<q_b$, which are treated in a "continuous" way, the following integral 
has to be evaluated:
%%%%%%%%%%%%%%%%%%%%%%%%%%%%%%%%%%%%%%%%%%%%%%%
\begin{equation}
\int\limits_0^{q_b} \left [1-F_a(q)\right ]^2\frac{{\d} q}{q}.       %12
\end{equation}
%%%%%%%%%%%%%%%%%%%%%%%%%%%%%%%%%%%%%%%%%%%%%%
A similar integral comes into cross section formulae for many electromagnetic 
interaction processes (for example, radiation logarithm in the 
bremsstrahlung), and a ready result of work [11] can be used. In the frame 
of the Thomas-Fermi model, one can find for the "restricted" value of 
squared transverse momentum accumulated in the layer $X$ of matter:
%%%%%%%%%%%%%%%%%%%%%%%%%%%%%%%%%%%%%%%%%%%%%%%
\begin{equation}
\langle Q^2\rangle_{\mathrm{rstr}}=8\pi N_{Av}(r_em_e)^2\frac{Z^2}{A}X\left 
[\ln KZ^{-1/3}+\ln\frac{q_b}{m_e}-1\right ],                %13
\end{equation}
%%%%%%%%%%%%%%%%%%%%%%%%%%%%%%%%%%%%%%%%%%%%%%
where $K$=183. Using Eq.(8) and Eq.(9), we can re-write the latter 
relation as
%%%%%%%%%%%%%%%%%%%%%%%%%%%%%%%%%%%%%%%%%%%%%%%
\begin{equation}
\langle Q^2\rangle_{\mathrm{rstr}}=2q_c^2\ln\frac{q_b}{q_a}.          %14
\end{equation}
%%%%%%%%%%%%%%%%%%%%%%%%%%%%%%%%%%%%%%%%%%%%%%
For light and medium elements, Thomas-Fermi model of the atom does not 
provide high accuracy. For example, for hydrogen it is necessary to take 
$K=$202.4; the values of $K$ for various elements calculated with 
Hartree-Fock model may be found elsewhere [12].\\
\hspace*{0.6cm}
Random scatters with $q>q_b$ are simulated by means of a usual technique, 
with the mean free path defined by the integral cross section (3). 
Random $q_{\varsigma}$ are sampled as
%%%%%%%%%%%%%%%%%%%%%%%%%%%%%%%%%%%%%%%%%%%%%%%
\begin{equation}
q_{\varsigma}=q_b/\sqrt{\varsigma},                                     %15
\end{equation}
%%%%%%%%%%%%%%%%%%%%%%%%%%%%%%%%%%%%%%%%%%%%%%
where $\varsigma$ is a random number with a uniform distribution between 0 
and 1; $q_{\varsigma}$ is accepted with a probability equal to the combined 
nuclear formfactor ${\cal {F}}_N^2$ (including elastic and quasielastic terms):
%%%%%%%%%%%%%%%%%%%%%%%%%%%%%%%%%%%%%%%%%%%%%%%
\begin{equation}
P(q_{\varsigma})={\cal {F}}_N^2 =F_N^2+\frac{1}{Z}\left (1-F_N^2\right )
F_p^2.                                                                 %16
\end{equation}
%%%%%%%%%%%%%%%%%%%%%%%%%%%%%%%%%%%%%%%%%%%%%%
Again, similar to Eq.(4), only the first term remains for hydrogen (with 
nuclear formfactor equal to that of proton).\\
\hspace*{0.6cm}
Although the consistency of the approach based on the partition of individual 
scatters into "continuous" ($q<q_b$) and "discrete" ($q>q_b$) parts is almost 
obvious, we have performed comparative calculations for different thicknesses 
of absorber (1 cm and 100 cm iron), each with two substantially different 
values
 of $q_b$ (0.1 and 1 MeV for thin absorber, 1 and 10 MeV for thick layer); 
$10^6$ muons for the lower value of $q_b$ and $10^8$ for the higher one 
have been traced. Noteworthy, the number of simulated individually scatters 
(per muon) varies as $1/q_b^2$  (see Eq.(3)), and hence it was 100 times 
different (about 15 per layer for the higher value of the threshold and 
$\sim$ 1500 for the lower one).\\
\hspace*{0.6cm}
Differential distributions of events in the value of accumulated transverse 
momentum $Q$ are shown in Figs.1 and 2 for 1 cm and 100 cm iron targets, 
respectively. A big difference between calculation results for point-like and 
finite-size nucleus is obvious already for 1 cm layer. One the other hand, the 
results obtained with different boundary values $q_b$ agree within the 
statistical accuracy in the whole range of angular deviations.\\
\hspace*{0.6cm}
Monte Carlo approach allows to easily probe different models of the nuclear 
formfactor. In Fig.3, comparison of calculation results obtained for several 
versions of the formfactor is given. Integral distributions in $Q$ (i.e. 
 probabilities of the deflection at angles greater than $Q/E$) after 
passing 100 cm iron layer are presented. Solid curves (bottom to top) in 
the figure correspond to: Gaussian charge distribution in nucleus, incoherent 
scattering on protons being neglected (i.e., the second term in Eq.(16) is 
omitted); Gaussian charge distribution both in nucleus and in proton; 
exponential distribution for nucleus and proton; exponential model for 
nucleus with proton taken as a point-like object ($F_p$=1). With the 
exception of unrealistic extreme cases, calculation results are not very 
sensitive to the choice of a specific formfactor model. At the same time, 
the difference with the calculations performed for a point-like nucleus 
(dashed curve in the figure) is very big, and reaches the orders of magnitude
 in the probability of large deflections.\\
\hspace*{0.6cm}
The total average value of squared transverse momentum accumulated in the 
layer can be obtained, if we add the contribution of large deflections 
($q>q_b$) to the restricted value given by Eq.(14). For Gaussian charge 
distribution, such evaluation can be easily performed analytically:
%%%%%%%%%%%%%%%%%%%%%%%%%%%%%%%%%%%%%%%%%%%%%%%
\begin{equation}
\langle Q^2\rangle=2q_c^2\left 
[\ln\frac{1.2978q_N}{q_a}+\frac{1}{2Z}\ln \left(1+\frac{q_p^2}{q_N^2} \right)
 \right ],                                                      %17
\end{equation}
where $q_N$, $q_p$, $q_c$, and $q_a$ are defined by Eqs.(7)-(9). The 
average contribution of quasielastic processes (the second term in brackets)
 is small even for light elements; for example, in carbon the rms-deviation
increases by only 1\%. However, these interactions have to be taken into
account for adequate description of the distribution function in the tail 
region (see Fig.3).\\
\hspace*{0.6cm}
For three-dimensional calculations, when both angular deviation and lateral 
displacement of muon are of interest, the restricted contribution 
(of $q<q_b$) can also be calculated in a Gaussian approximation, but taking 
into account the well-known correlation between the angle and the displacement
 [13].
%_______________________________________________________
\section{Atomic electron contribution; mixtures and compounds}
\label{sect4}
\hspace*{0.6cm}
Contribution of incoherent scattering on atomic electrons is often taken into 
account by means of a substitution of $Z^2$ by $Z(Z+1)$ in basic relations 
(for differential cross section, some parameters of the theory, etc.). 
However, this substitution is rather inaccurate, since, firstly, the cross 
section at small angles (and low momenta) is determined in this case by the 
inelastic atomic formfactor (with different $q$-dependence and different 
characteristic radius of the atom). Secondly, due to light electron mass, 
kinematics of the process (in laboratory frame) appreciably changes. 
That leads to the appearance of the upper limit on transverse momentum and 
on deflection angle, and to an appreciable probability to loose large 
fraction of muon energy in a single collision. Of course, target size 
corrections for scattering on electrons are irrelevant.\\
\hspace*{0.6cm}
Consideration of kinematics of muon scattering on a free electron gives the 
following expression for the transverse momentum $q$
%%%%%%%%%%%%%%%%%%%%%%%%%%%%%%%%%%%%%%%%%%%%%%%
\begin{equation}
q=\sqrt{2m_eT(1-T/T_m)},                                  %18
\end{equation}
%%%%%%%%%%%%%%%%%%%%%%%%%%%%%%%%%%%%%%%%%%%%%%
where $T$ and $T_m$ are kinetic energy transferred to the electron (initially 
at rest) in the collision and the maximal value of this energy:
%%%%%%%%%%%%%%%%%%%%%%%%%%%%%%%%%%%%%%%%%%%%%%%
\begin{equation}
T_m=2m_ep^2/(m_e^2+m_{\mu}^2+2m_eE).                       %19
\end{equation}
%%%%%%%%%%%%%%%%%%%%%%%%%%%%%%%%%%%%%%%%%%%%%%
The accumulated squared transverse momentum in the layer $X$ may be evaluated 
similar to Eq.(13), combining the contribution of low-$q$ region (where the 
inelastic atomic formfactor is important) and of high-$q$ domain. In the 
latter case, $\langle Q^2\rangle$ is easily calculated by means of the 
convolution of the squared value of $q$ ( Eq.(18)) and the 
differential cross section of knock-on electron production. In this way, 
we obtain the following relation:
%%%%%%%%%%%%%%%%%%%%%%%%%%%%%%%%%%%%%%%%%%%%%%%
\begin{equation}
\langle Q^2 \rangle=8\pi N_{\mathrm{Av}}\frac{Z}{A}X(r_em_e)^2\left[\ln\left 
(K^{\prime} Z^{-2/3}\sqrt{\frac{2T_m}{m_e}}\right)-\frac{7}{4}\right].   %20
\end{equation}
%%%%%%%%%%%%%%%%%%%%%%%%%%%%%%%%%%%%%%%%%%%%%%
Within the Thomas-Fermi model, $K^{\prime}$=1429; an accurate value 
for hydrogen is $K^{\prime}$=446. Small corrections which correspond to spin 
terms in the cross section of muon-electron scattering are omitted for 
simplicity. \\
\hspace*{0.6cm}
Rare catastrophic collisions with large energy transfers are 
also included in $\langle Q^2 \rangle$ calculations (Eq.(20)) on the average. 
For simulation of muon transport in thick layers, such collisions are usually 
treated  separately (with explicit simulation of kinematic variables). 
The average accumulated $Q^2$ in collisions with energy transfers less than 
a certain cut $T_c$ ($T_c\le T_m$) is given by
%%%%%%%%%%%%%%%%%%%%%%%%%%%%%%%%%%%%%%%%%%%%%%%
\begin{equation}
\langle Q^2 \rangle_{\mathrm{rstr}}=8\pi N_{\mathrm{Av}}\frac{Z}{A}X(r_em_e)^2
\left[\ln\left (K^{\prime} Z^{-2/3}\sqrt{\frac{2T_c}{m_e}}\right)-1-
\frac{T_c}{T_m}+\frac{T_c^2}{4T_m^2}\right].                           %21
\end{equation}
%%%%%%%%%%%%%%%%%%%%%%%%%%%%%%%%%%%%%%%%%%%%%%
Multiple scattering on the electrons represents only a correction 
to the scattering on nuclei, and maximal transverse momenta and deflection 
angles are small. Therefore it seems justified to introduce this correction 
on the basis of a Gaussian approximation (for example, as an addition of 
contribution given by Eq.(21) to the restricted value of accumulated $Q^2$ on 
nuclei defined by Eq.(13)). Such procedure allows to avoid double counting of 
the collisions with energy transfers $T>T_c$, which will appear if multiple 
scattering process and knock-on electron production are simulated 
incoherently.\\
\hspace*{0.6cm}
As a general prescription, for calculation of multiple scattering in mixtures
 or compounds, the differential cross sections Eq.(4) for elements have to be 
averaged with weights $W_i$ corresponding to relative mass abundance of 
individual entries.\\
\hspace*{0.6cm}
Within the framework of a mixed Monte Carlo approach described above, 
this leads 
to: (i) summation of restricted accumulated $Q^2$-values with weights $W_i$; 
(ii) description of the cross section in the intermediate region ($q_a\ll q
\ll q_N$) with the average parameter $\langle Z^2/A\rangle $ (and, hence, of 
the integral cross section for $q>q_b$ with the same factor); (iii) averaging 
of the acceptance function Eq.(16) for sampling of random momenta 
$q_{\varsigma}
>q_b$:
%%%%%%%%%%%%%%%%%%%%%%%%%%%%%%%%%%%%%%%%%%%%%%%
\begin{equation}
P(q_{\varsigma})=\left.\sum_{i=1}^n W_i\frac{Z_i^2}{A_i}\left \{F_{Ni}^2+
\frac{1}{Z_i}
(1-F_{Ni}^2)F_p^2\right \}\right /\sum_{i=1}^n W_i\frac{Z_i^2}{A_i}.       %22
\end{equation}
%%%%%%%%%%%%%%%%%%%%%%%%%%%%%%%%%%%%%%%%%%%%%%
Here, $F_{Ni}(q_{\varsigma})$ denotes the elastic nuclear formfactor for the 
$i$-th element of the mixture. For hydrogen, the expression in braces reduces 
to the first term (see also comments after Eqs.(4) and (16)). Similarly, 
the contribution of the scattering on atomic electrons is 
determined by the weighted sum of partial contributions defined by Eqs.(20) or
 (21), for total or restricted (with $T< T_c$) accumulated $Q^2$ values, 
respectively.
%____________________________________________
\section{Moliere's equations}
\label{sect5}
\hspace*{0.6cm}
In this section, we give the main equations of the Moliere's multiple 
scattering theory where the nucleus is treated as a point-like charge.\\
\hspace*{0.6cm}
Let $f(\theta,t)$ be spatial-angle distribution function, i.e. for small-angle
 approximation $f(\theta,t)$ it is the number of charged particles in the 
angular interval of polar angle $\theta\div \theta +{\d} \theta$ after 
traversing a thickness of $t$.   
Moliere solved the transport equation for determination of
$f(\theta,t)$ and obtained the general expression for angular distribution
function as a following:
%%%%%%%%%%%%%%%%%%%%%%%%%%%%%%%%%%%%%%%%%%%%%%%%%
\begin{equation}
f(\theta,t)=\frac{1}{2 \pi}\int\limits_0^\infty \xi {\d}\xi J_0(\xi \theta)
\e^{\Omega(\xi,t)},                                                 %23
\end{equation}
where $J_0(\xi \theta)$ is the Bessel function of order 0.
%%%%%%%%%%%%%%%%%%%%%%%%%%%%%%%%%%%%%%%%%%%%%%%%%
For a homogeneous scatterer without energy loss, we have
%%%%%%%%%%%%%%%%%%%%%%%%%%%%%%%%%%%%%%%%%%%%%%%%
\begin{equation}
\Omega(\xi)=2\pi t \int\limits_0^\infty\chi {\d}\chi [J_0(\xi \chi)-1]
W(\chi),                                                        %24
\end{equation}
%%%%%%%%%%%%%%%%%%%%%%%%%%%%%%%%%%%%%%%%%%%%%%%%
where
%%%%%%%%%%%%%%%%%%%%%%%%%%%%%%%%%%%%%%%%%%%%%%%%%%%%%%%%%%%%%%%
\begin{equation}
W(\chi)=N\sigma(\chi).                                    %25
\end{equation}
%%%%%%%%%%%%%%%%%%%%%%%%%%%%%%%%%%%%%%%%%%%%%%%%%%%%%%%%%%%%%%%
Here $N=N_{Av}\rho /A$ ($\rho$ is density of the scatterer in g/cm$^3$) 
is the number of scattering atoms per unit volume, with 
differential cross-section $\sigma(\chi)$ each.\\
\hspace*{0.6cm}   
The scattering of relativistic particle by an atom is determined by a 
modified Rutherford law 
%%%%%%%%%%%%%%%%%%%%%%%%%%%%%%%%%%%%%%%%%%%%%%%%%
\begin{equation}
\sigma(\chi)=\sigma_{\mathrm{Ru}}(\chi)\Psi(\chi){\cal F}_N^2(\chi),  %26
\end{equation}
where $\sigma_{\mathrm{Ru}}(\chi)$ is Rutherford cross-section for scattering 
by point-like charge, $\Psi(\chi)$ is a function which takes into account the 
screening of the nuclear Coulomb field by atomic electrons, and 
$\cal {F}_N(\chi)$ is the nuclear formfactor Eq.(16). In Moliere's theory, 
the following approximations  are used: $ {\cal F}_N^2(\chi)$=1,
%%%%%%%%%%%%%%%%%%%%%%%%%%%%%%%%%
\begin{equation}
\sigma_{\mathrm{Ru}}(\chi)=4\frac{Z^2e^4}{p^2\chi^4},   %27
\end{equation}
%%%%%%%%%%%%%%%%%%%%%%%%%%%%%%%%%
and the screening function is 
%%%%%%%%%%%%%%%%%%%%%%%%%%%%%%%%%
\begin{equation}
\Psi(\chi)=\chi^4/(\chi^2+\chi_\alpha ^2)^2,     %28
\end{equation}
%%%%%%%%%%%%%%%%%%%%%%%%%%%%%%%%%
where the screening angle is
%%%%%%%%%%%%%%%%%%%%%%%%%%%%%%%%%%
\begin{equation}
\chi^2_{\alpha}=\chi^2_0\left (1.13+3.76(Z \alpha)^2\right ),           %29  
\end{equation}
%%%%%%%%%%%%%%%%%%%%%%%%%%%%%%%%%%%%%%%%%%%%%
\[
\alpha=1/137,~~~~~\mbox{and}~~~~~\chi_0\simeq\frac{1.13}{137}Z^{1/3}m_e/p.
\] 
%%%%%%%%%%%%%%%%%%%%%%%%%%%%%%%%%%%%%%%%%%%%%
\hspace*{0.6cm}
In his original paper, Moliere used the variables
\begin{equation}
\vartheta=\theta/\chi_c\sqrt{B}~~~~\mbox{and}~~~~\eta=\xi\chi_c\sqrt{B}, %30
\end{equation}
%%%%%%%%%%%%%%%%%%%%%%%%%%%%%%%%%%%%%%%%%%%%%%%%%%%%%%%
where the characteristic angle $\chi_c$ is defined as
%%%%%%%%%%%%%%%%%%%%%%%%%%%%%%%%%%%%%%%%%%%%%%%%%%%%%%%%%%%%%
\begin{equation}
\chi^2_c=\frac{4\pi NtZ^2e^4}{p^2}.                      %31
\end{equation}
%%%%%%%%%%%%%%%%%%%%%%%%%%%%%%%%%%%%%%%%%%%%%%%%%%%%%%%%%%%%%%
Here $t$ is measured in cm, and $B$ equals to the solution of the 
transcendental equation
%%%%%%%%%%%%%%%%%%%%%%%%%%%%%%%%%%%%%%%%%%%%%%%%%%%%%%%%%
\begin{equation}
B-\ln B=1-2C+\ln\left (\chi_c^2/\chi_\alpha^2\right )       %32      
\end{equation}
%%%%%%%%%%%%%%%%%%%%%%%%%%%%%%%%%%%%%%%%%%%%%%%%%%%%%%%%%
($C=0.5772$ is Eiler constant). Then the expression for the angular 
distribution function can be written as
%%%%%%%%%%%%%%%%%%%%%%%%%%%%%%%%%%%%%%%%%%%%%%%%%%%%%
\begin{equation}
f(\theta,t)\theta {\d}\theta=f_M(\vartheta,B)\vartheta \d\vartheta,    %33
\end{equation}
where
\[
f_M(\vartheta,B)=\int\limits_0^\infty \eta {\d}\eta J_0(\eta \vartheta)
\e^{-\eta^2/4}\exp\left (\frac{\eta^2}{4B}\ln\frac{\eta^2}{4}\right ).   
\]
Moliere's expansion method is to consider the term $\left [\eta^2\
ln(\eta^2/4)\right ]/4B$
as a small parameter; then the exponent can be expanded to second order terms
%%%%%%%%%%%%%%%%%%%%%%%%%%%%%%%%%%%%%%%%%%%%%%%%%%%%%
\begin{equation}
f_M(\vartheta,B)=f_M^{(0)}(\vartheta)+f_M^{(1)}(\vartheta)/B+f_M^{(2)}
(\vartheta)/B^2+ \cdots,                                           %34
\end{equation}
where
\begin{equation}
f_M^{(n)}(\vartheta)=\frac{1}{n!}\int\limits_0^\infty \eta {\d}\eta 
J_0(\vartheta\eta)\e^{-\eta^2/4}\left [\frac{\eta^2}{4}\ln\frac{\eta^2}{4}
\right ]^n.                                                         %35
\end{equation}
%%%%%%%%%%%%%%%%%%%%%%%%%%%%%%%%%%%%%%%%%%%%%%%%%%%%%%%
The first two functions $f^{(n)}$ have simple analytical forms:
%%%%%%%%%%%%%%%%%%%%%%%%%%%%%%%%%%%%%%%%%%%%%%%%%%%%%%
\begin{equation}
f_M^{(0)}(\vartheta)=2\e^{-\vartheta^2},~~~~~~~  f_M^{(1)}(\vartheta)=
2\e^{-\vartheta^2}\left (\vartheta^2-1\right )\left [E_1(\vartheta^2)-
\ln\vartheta^2\right ]-
2\left (1-2\e^{-\vartheta^2}\right ),                            %36
\end{equation}
%%%%%%%%%%%%%%%%%%%%%%%%%%%%%%%%%%%%%%%%%%%%%%%%%%%%
where $E_1\left (\vartheta^2\right )$ is the exponential integral. 
The Moliere's theory is valid for $B\ge 4.5$ and $\chi_c^2B<1$ [2],[3].
%____________________________________________________
\begin{section}
{Modified Moliere theory}  
\end{section}
\label{sect6}
\hspace*{0.6cm}
The effect of nuclear size can  be evaluated using a suitable formfactor
${\cal F}_N(\chi)$ in Eq.(26). The general method to take into account
nuclear effects for arbitrary charge distribution in nucleus has been 
suggested by Cooper and Rainwater [4]. The Gaussian approximation of 
nucleus and protons formfactors was regarded in paper [5]. In this case, 
assuming that $q_p \gg q_N$ one can re-write the Eq.(16) as
%%%%%%%%%%%%%%%%%%%%%%%%%%%%%%%%%%%%%%%%%%%%%%%%%%%%%%%%%%%%%%%%%
\begin{equation}
{\cal {F}}^2_N(\chi)=\left (1-Z^{-1}\right )\e^{-\chi^2/a_A^2}+Z^{-1}
\e^{-\chi^2/a_p^2},                                                 %37
\end{equation}
%%%%%%%%%%%%%%%%%%%%%%%%%%%%%%%%%%%%%%%%%%%%%%%%%%%%%%%%%%%%%
where (see Eq.(7) and Eq.(16))
%%%%%%%%%%%%%%%%%%%%%%%%%%%%%%%%%%%%%%%%%%%%%%%%%%%%%%%%%%%%%%%%%
\[
a_A^2={3q_N^2}/{E^2}~~~~~~\mbox{and}~~~~~~a_p^2={3q_p^2}/{E^2}. 
\]
%%%%%%%%%%%%%%%%%%%%%%%%%%%%%%%%%%%%%%%%%%%%%%%%%%%%%%%%%%%%%
Using Eqs.(24)-(28) and Eq.(37), the function $\Omega(\xi)$ can 
be written as
%%%%%%%%%%%%%%%%%%%%%%%%%%%%%%%%%%%%%%%%%%%%%%%%%%%%
\begin{equation}
\Omega(\xi)=2\chi_c^2\int\limits_0^\infty \frac{\chi {\d}\chi}{\left (\chi^2+
\chi_\alpha^2\right )^2}
\left [J_0(\xi\chi)-1\right ]\left [\left (1-Z^{-1}\right )\e^{-\chi^2/a_A^2}
+Z^{-1}\e^{-\chi^2/a_p^2}\right ] .                             %38
\end{equation}
%%%%%%%%%%%%%%%%%%%%%%%%%%%%%%%%%%%%%%%%%%%%%%%%%%%%%%%%%%%%%
This integral can be evaluated analytically, and we have
%%%%%%%%%%%%%%%%%%%%%%%%%%%%%%%%%%%%%%%%%%%%%%%%%%%%%%%%%%%%%
\begin{equation}
\Omega(\xi)=-\frac{\eta^2}{4}+\frac{1}{B}\left [\left (1-Z^{-1}\right )
D_A\left (\eta,\tau^2_A\right )+Z^{-1}D_p\left (\eta,\tau_p^2\right )\right ],
                                                               %39
\end{equation}
%%%%%%%%%%%%%%%%%%%%%%%%%%%%%%%%%%%%%%%%%%%%%%%%%%%%%%%%%%%%%%%
where
%%%%%%%%%%%%%%%%%%%%%%%%%%%%%%%%%%%%%%%%%%%%%%%%%%%%%%%%%%%%%%%
\begin{equation}
D\left (\eta,\tau^2\right )=\frac{\eta^2}{4}\left [\ln\frac{\eta^2}{4}
+E_1\left (\frac{\eta^2\tau^2}{4}\right )\right ]+\frac{1}{\tau^2}
\left [C+\ln\frac{\tau^2\eta^2}{4}+1-\e^{-\eta^2\tau^2/4}+E_1\left 
(\frac{\eta^2\tau^2}{4}\right )\right ],  %40    
\end{equation}
%%%%%%%%%%%%%%%%%%%%%%%%%%%%%%%%%%%%%%%%%%%%%%%%%%%%%%%%%%%%%%%
and
\[
\tau^2=a^2/\chi_c^2B.
\]
Then angular distribution function $\tilde f_M(\vartheta,\tau)$ is
%%%%%%%%%%%%%%%%%%%%%%%%%%%%%%%%%%%%%%%%%%%%%%%%%%%%%%%%%%%%%%%%%%
\begin{equation}
\tilde f_M(\vartheta,\tau)=\int\limits_0^\infty \eta {\d}\eta 
J_0(\eta,\vartheta)
\e^{-\eta^2/4}\exp\left \{\frac{1}{B}\left [\left (1-Z^{-1}\right )
D\left (\eta,\tau^2_A\right )+Z^{-1}D\left (\eta,\tau^2_p\right )\right ]
\right \}.                                                        %41
\end{equation}
%%%%%%%%%%%%%%%%%%%%%%%%%%%%%%%%%%%%%%%%%%%%%%%%%%%%%%%%%%%%%%%
The exponent in Eq.(41) may be expanded in a series, and we obtain
%%%%%%%%%%%%%%%%%%%%%%%%%%%%%%%%%%%%%%%%%%%%%%%%%%%%%%%%%%%%%%%%
\begin{equation}
\tilde f_M(\vartheta,\tau)=\sum_{n=0}^\infty \frac{1}{B^n}\tilde f_M^{(n)}
(\vartheta,\tau)\approx \tilde f_M^{(0)}(\vartheta)+\frac{1}{B}\tilde 
f_M^{(1)}(\vartheta,\tau)+\cdots,                                   %42
\end{equation}
%%%%%%%%%%%%%%%%%%%%%%%%%%%%%%%%%%%%%%%%%%%%%%%%%%%%%%%%%%%%%
where
\begin{equation}
\tilde f_M^{(n)}(\vartheta,\tau)=\frac{1}{n!}\int\limits_0^\infty \eta{\d}\eta 
J_0(\eta,\vartheta)\e^{-\eta^2/4}\left [\left (1-Z^{-1}\right )
D\left (\eta,\tau^2_A\right )+Z^{-1}D\left (\eta,\tau^2_p\right )\right ]^n.%43
\end{equation}
%%%%%%%%%%%%%%%%%%%%%%%%%%%%%%%%%%%%%%%%%%%%%%%%%%%%%%%%%%%%%%
At $n=0$ and $1$, the integrals Eq.(43) equal to
\footnote{Note that integrals with $E_1\left (\eta^2\tau^2/4\right )$ can be 
evaluated using derivation over the parameter $\tau^2$.}
%%%%%%%%%%%%%%%%%%%%%%%%%%%%%%%%%%%%%%%%%%%%%%%%%%%%%%%%%%%%%%
\begin{equation}
\tilde f_M^{(0)}(\vartheta)=2\e^{-\vartheta^2}~,~~~~\tilde f_M^{(1)}(\vartheta,
\tau)=2\e^{-\vartheta^2}\left [2+\left (1-Z^{-1}\right )G\left (\vartheta,
\tau^2_A\right )+Z^{-1}G\left (\vartheta,\tau_p^2\right )\right ],     %44
\end{equation}
%%%%%%%%%%%%%%%%%%%%%%%%%%%%%%%%%%%%%%%%%%%%%%%%%%%%%%%%%%%%
where
%%%%%%%%%%%%%%%%%%%%%%%%%%%%%%%%%%%%%%%%%%%%%%%%%%%%%%%%%%%
\begin{equation}
G\left (\vartheta,\tau^2\right )=\frac{1}{\tau^2}\left (C+1+\ln\tau^2\right )+
\kappa^2\left (\frac{\vartheta^2}
{\kappa^2}-1\right )\left [Ei(\frac{\vartheta^2}{\kappa^2})-\ln\vartheta^2
\right ]-\kappa^2
\e^{\vartheta^2/\kappa^2},                                         %45
\end{equation}
%%%%%%%%%%%%%%%%%%%%%%%%%%%%%%%%%%%%%%%%%%%%%%%%%%%%%%%%%%%%%
and $\kappa^2=1+1/\tau^2$ (in Ref.[5] the coefficient $\kappa^2$ is 
missing in the last term).\\
\hspace*{0.6cm}
The integral angular distribution $P(\vartheta,t)$ is
%%%%%%%%%%%%%%%%%%%%%%%%%%%%%%%%%%%%%%%%%%%%%%%%%%%%%%%%%%%%
\begin{equation}
P(\vartheta,t)=\sum_{n=0}^\infty \frac{1}{B^n}\int\limits_\vartheta^\infty 
\tilde f_M^{(n)}(\vartheta ',\tau)\vartheta' {\d}\vartheta' \approx     %46
P^{(0)}(\vartheta)+P^{(1)}(\vartheta,\tau)+\cdots, 
\end{equation}
%%%%%%%%%%%%%%%%%%%%%%%%%%%%%%%%%%%%%%%%%%%%%%%%%%%%%%%%%%%%%%%
where
\begin{equation}
P^{(0)}(\vartheta)=\e^{-\vartheta^2}~,~~~~P^{(1)}(\vartheta,\tau)=
\frac{1}{B}\left [\left (1-Z^{-1}\right )Q\left (\vartheta,\tau^2_A\right )+
Z^{-1}Q\left (\vartheta,\tau^2_p\right )\right ],                  %47
\end{equation}
%%%%%%%%%%%%%%%%%%%%%%%%%%%%%%%%%%%%%%%%%%%%%%%%%%%%%%%%%%%%%%%%
and 
%%%%%%%%%%%%%%%%%%%%%%%%%%%%%%%%%%%%%%%%%%%%%%%%%%%%%%%%%%%%%%%%
\begin{equation}
Q(\vartheta,\tau^2)=\e^{-\vartheta^2}\left \{\frac{\tau^2+C+1+\ln\tau^2}
{\tau^2}+(\kappa^2-1)\left [R_1\left (\vartheta,\tau^2\right )+R_2\left 
(\vartheta,\tau^2\right )\right ]\right \}.                        %48
\end{equation}
%%%%%%%%%%%%%%%%%%%%%%%%%%%%%%%%%%%%%%%%%%%%%%%%%%%%%%%%%%%%%%%%%
The functions $R_1$ and $R_2$ are
\[
R_1\left (\vartheta,\tau^2\right )=\e^{\vartheta^2}\left 
[Ei\left (-\frac{\vartheta^2}       
{1+\tau^2}\right )-Ei(-\vartheta^2)\right ],
\]
%%%%%%%%%%%%%%%%%%%%%%%%%%%%%%%%%%%%%%%%%%%%%%%%%%%%%%%%%%%%%%%%
\[
R_2\left (\vartheta,\tau^2\right )=\left (\frac{\vartheta^2}
{\kappa^2-1}-1\right )\left [Ei\left (\frac{\vartheta^2}
{\kappa^2}\right )-\ln\vartheta^2\right ]-\kappa^2\e^{\vartheta^2/\kappa^2}.
\]
%%%%%%%%%%%%%%%%%%%%%%%%%%%%%%%%%%%%%%%%%%%%%%%%%%%%%%%%%%%%%%%%%
\hspace*{0.6cm}
The function $\tilde f_M(\vartheta,\tau) \to$ $f_M(\vartheta)$
at $\langle R^2_N\rangle$ and $\langle R^2_p\rangle \to 0$. On the other 
hand, $\tau^2 \to \infty$ at $t\to 0$ also. So, the angular distribution 
of relativistic particles should be similar to Moliere distribution after 
passing through the small thicknesses of matter. However, 
 even for thin layers the distribution function appreciably
deviates from the Moliere law at large deflection angles ($\theta > q_N/E$), 
and tends to the cross section behaviour Eq.(26) with formfactors. 
The expansions to the first 
order terms of Eq.(42) and Eq.(46) are valid up to $\tau^2_A\ge 0.2$. For
 thick layer of matter ($\tau^2_A$ and $\tau^2_p\ll 1$), the 
assymptotic of Eq.(42) is Gaussian [5]: 
%%%%%%%%%%%%%%%%%%%%%%%%%%%%%%%%%%%%%%%%%%%%%%%%%%%%%%%%%%%%%
\begin{equation}
\tilde f_M(\vartheta,\tau)\simeq\int\limits_0^\infty J_0(\eta\vartheta)
\e^{-\mu\eta^2/4}\eta {\d}\eta=\frac{2}{\mu}\e^{-\vartheta^2/\mu},       %49
\end{equation}
%%%%%%%%%%%%%%%%%%%%%%%%%%%%%%%%%%%%%%%%%%%%%%%%%%%%%%%%%%%%%%%
where
\begin{equation}
\mu=1-\frac{1}{B}\left [2-C-\ln\tau^2_A-Z^{-1}\ln\frac{\tau^2_p}{\tau^2_A}
\right ].
\end{equation}                                                        %50 
%%%%%%%%%%%%%%%%%%%%%%%%%%%%%%%%%%%%%%%%%%%%%%%%%%%%%%%%%%%%%%%%%%%%
\hspace*{0.6cm}
If the mixture of $n$ different scatterers is present, the characteristic 
angle $\chi_c$ is given by
%%%%%%%%%%%%%%%%%%%%%%%%%%%%%%%%%%%%%%%%%%%%%%%%%%%%%%%%%%%%%
\begin{equation}
\chi^2_c=\frac{4\pi e^4}{p^2}N_{\mathrm{Av}}\rho t\left.\sum_{i=1}^n 
w_iZ_i^2\right/\sum_{i=1}^n w_iA_i,                                   %51
\end{equation}
%%%%%%%%%%%%%%%%%%%%%%%%%%%%%%%%%%%%%%%%%%%%%%%%%%%%%%%%%%%%%%
and the integral angular distribution function $P(\vartheta,t)$ is defined by
%%%%%%%%%%%%%%%%%%%%%%%%%%%%%%%%%%%%%%%%%%%%%%%%%%%%%%%%%%%%%
\begin{equation}
P(\vartheta,t)=\left.\sum_{i=1}^n w_iZ_i^2P_i(\vartheta,t)\right/\sum_{i=1}^n 
w_iZ_i^2,                                                           %52
\end{equation}
%%%%%%%%%%%%%%%%%%%%%%%%%%%%%%%%%%%%%%%%%%%%%%%%%%%%%%%%%%%%%%
where the subscript $i$ denotes the atomic species and weight $w_i$ 
is proportional to the number of atoms of the $i$-th element in the mixture.\\
\hspace*{0.6cm}
The integral angular distributions on nuclei as a function of the transverse 
momentum $Q$ calculated by the mixed Monte Carlo tecnhique and analytically 
Eq.(46) are shown in Fig.4 (iron targets of 1 cm and 100 cm) 
and in Fig.5 (water thickness of 1000 cm). The results are in good 
agreement. Moliere angular distributions are shown also for comparison. A 
minor difference between analytical and Monte Carlo calculations in the 
tails of the curves is conditioned by a different choice of the proton 
formfactor (Gaussian charge distribution with $R_p=0.80$ fm in analytical
evaluation and exponential one with $R_p=0.85$ fm in Monte Carlo simulation. 
%________________________________________________
\section{ Conclusions}
\label{sect7}
\hspace*{0.6cm}
A high-energy muon is capable to cross very thick layers of material and its
de Broglie wave is comparable with nuclear dimensions. Hence, the
probability of scatters with impact parameter of the order of nuclear radius
( scatters with large angles) is not small. For such interactions, the law of 
single scattering has to be modifed significantly to take into account 
the finite size of charge distribution in the nuclei. Nuclear size effects can be desribed by multiplying the Rutherford cross section by a nuclear 
formfactor ${\cal {F}}_N^2$, which goes to 1 for large impact parameters 
and vanished at very large transverse momenta.  \\
\hspace*{0.6cm}   
We have presented two methods for calculating the multiple scattering
distribution for muons traversing a thick 
scatterer. The mixed Monte Carlo technique and modified Moliere's theory of
MCS take into account the nuclear size effect.  
Parametrisation of $A$-dependence of 
rms-radii for light and medium nuclei is given. The results of two methods
are in good agreement, and show that the muon angular distribution is similar
to Moliere distribution after passing through small thickness of material and
moderate deflection angles.
However at large thickness of scatterer, the angular distribution is 
drastically changed by the influence of the nuclear formfactor. So, in this 
case the Moliere theory heavily overestimates the probability of large angle 
deflections. Contribution of incoherent scattering on atomic electrons is
also correctly taken into account.\\
\hspace*{0.6cm}
A necessity of taking into account nucleus size effect as well as incoherent 
scattering on atomic electrons when describing the muon scattering at large 
angles in thick matter layers  is shown. Neither Gaussian approximation 
of multiple scattering distribution nor Moliere theory can be used for accurate
description of high-energy muon transport in a wide range of the thicknesses
of material. 
%________________________________________________
\section*{Acknowledgements} 
\hspace*{0.6cm}
Authors are indebted to A.A. Petrukhin for stimulating interest and critical 
comments. We acknowledge helpful discussions with V.L. Matushko. The work 
was supported in part by Russian Federal Program "Integratsiya" (project 
A-0100) and RFBR grant 99-02-18374.

%%References start:
%%%%%%%%%%%%%%%%%%%%%%%%%%%%%%%%%%%%%%%%%%%%%%%%%%%%%%%%%%%%%%%%%%%%%%%%%%

%
\newpage
\section*{Figure Captions} 

Fig.1: Differential distribution in accumulated transverse momentum for 
muons after passing iron of 1 cm. Dashed curves, open circles: point-like 
nucleus. Solid curves, dark points: calculations taking into account the 
nuclear formfactor. Curves are calculated with $q_b$=1 MeV ($10^8$ events), 
the points correspond to $q_b$=0.1 MeV ($10^6$ events).

Fig.2: The same as in Fig.1 but for iron target of 100 cm. In this case, the 
values of $q_b$ in simulation were 10 MeV and 1 MeV, respectively.

Fig.3: Integral distributions in the value of accumulated transverse 
momentum for muons after passing iron layer of 100 cm, calculated with 
different models of the nucleus (see the text).

Fig.4: Integral distributions in the value of accumulated transverse 
momentum for muons after passing iron layers of 1 cm and 100 cm calculated by
the mixed Monte Carlo tecnique (open circles for  point-like nucleus and 
dark points for finite nuclei) and analytically (solid curves are result 
of Moliere theory and dashed curves are calculations taking into account
the nuclear size effect).  

Fig.5: Integral distributions in the value of accumulated transverse 
momentum for muons after passing water layer of 1000 cm calculated by
the mixed Monte Carlo tecnique (dark points for  point-like nucleus, 
open circles for finite nuclei and stars is a result of calculations, taking
into account the nuclear formfactor and electron contribution) and analytically
 (solid curves are result of Moliere theory and dashed curves are 
calculations taking into account the nuclear size effect).  

\begin{center}
\mbox{\epsfig{file=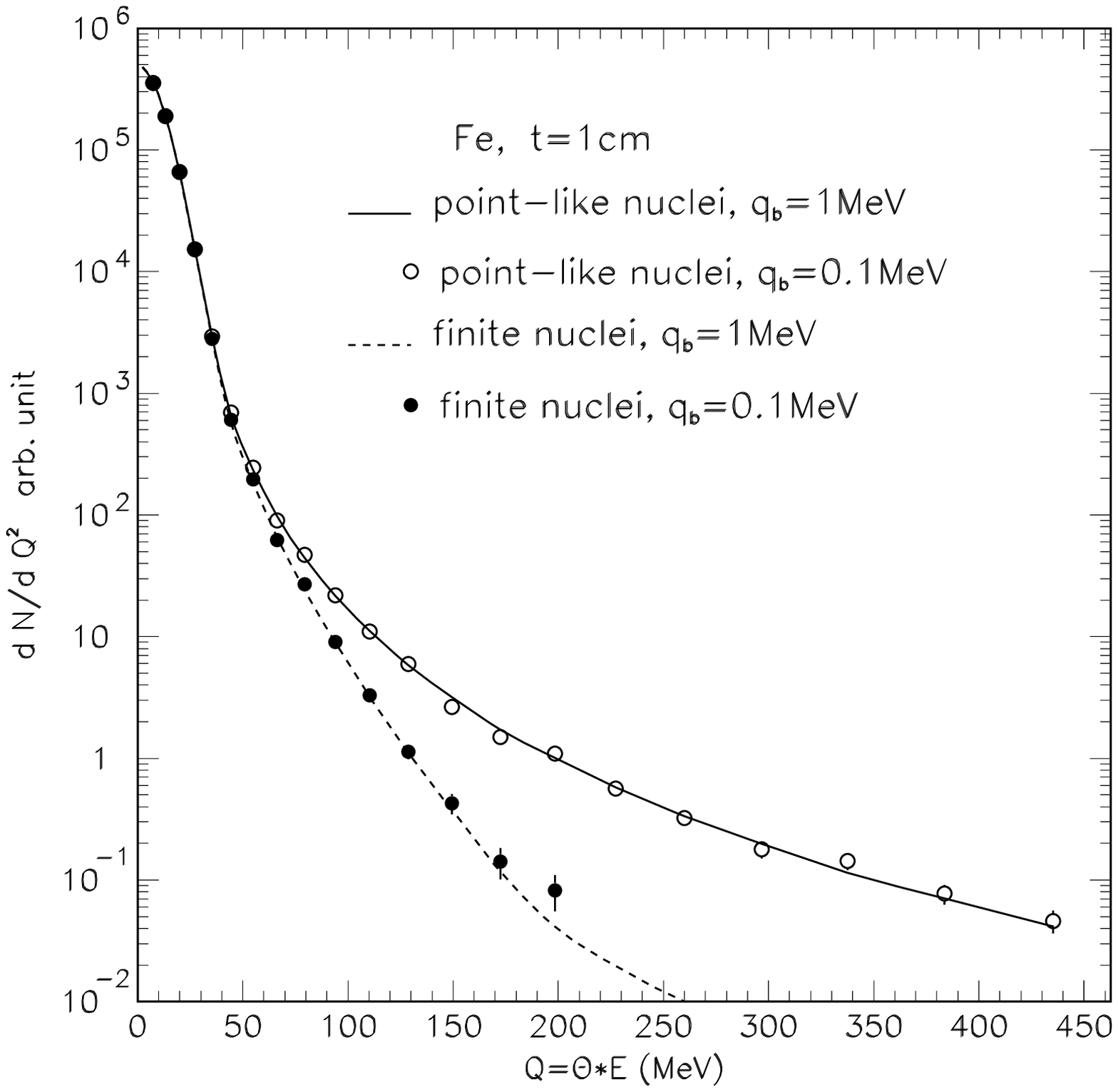,height=18cm,width=18cm}}

\vspace{2mm}
\noindent
\small
\end{center}
%%%%%%%%%%%%%%%%%%%%%%%%%%%%%%%%%
{\sf~~~~~~~~~~~~~~~~~~~~~~~~~~~~~~~~~~~~~~~~~~~~~~~~~~~~~~~ Fig.~1}
%%%%%%%%%%%%%%%%%%%%%%%%%%%%%%%%%
\begin{center}
\mbox{\epsfig{file=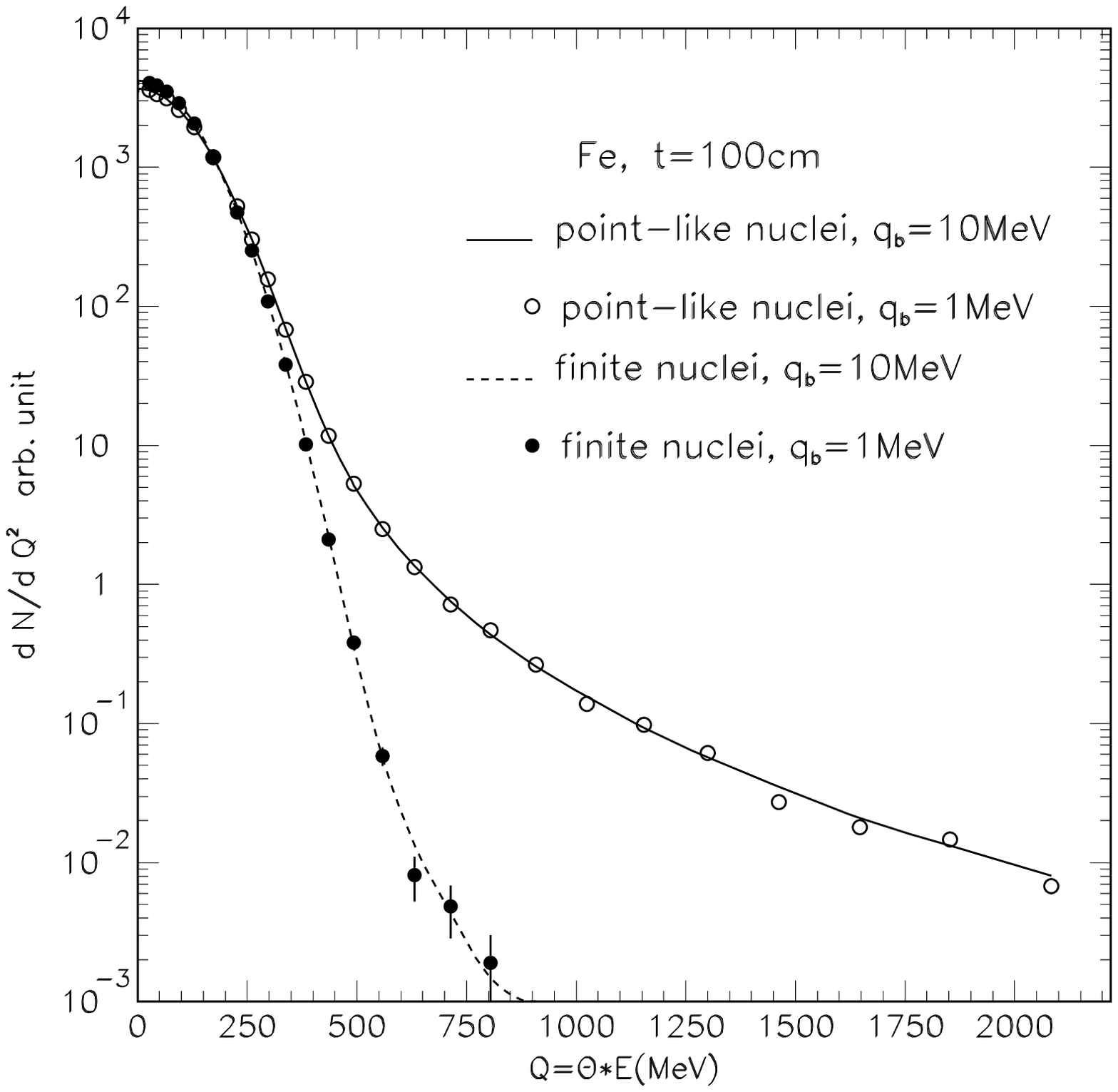,height=18cm,width=18cm}}

\vspace{2mm}
\noindent
\small
\end{center}
%%%%%%%%%%%%%%%%%%%%%%%%%%%%%%%%%
{\sf~~~~~~~~~~~~~~~~~~~~~~~~~~~~~~~~~~~~~~~~~~~~~~~~~~~~~~~~ Fig.~2}
%%%%%%%%%%%%%%%%%%%%%%%%%%%%%%%%%
\begin{center}
\mbox{\epsfig{file=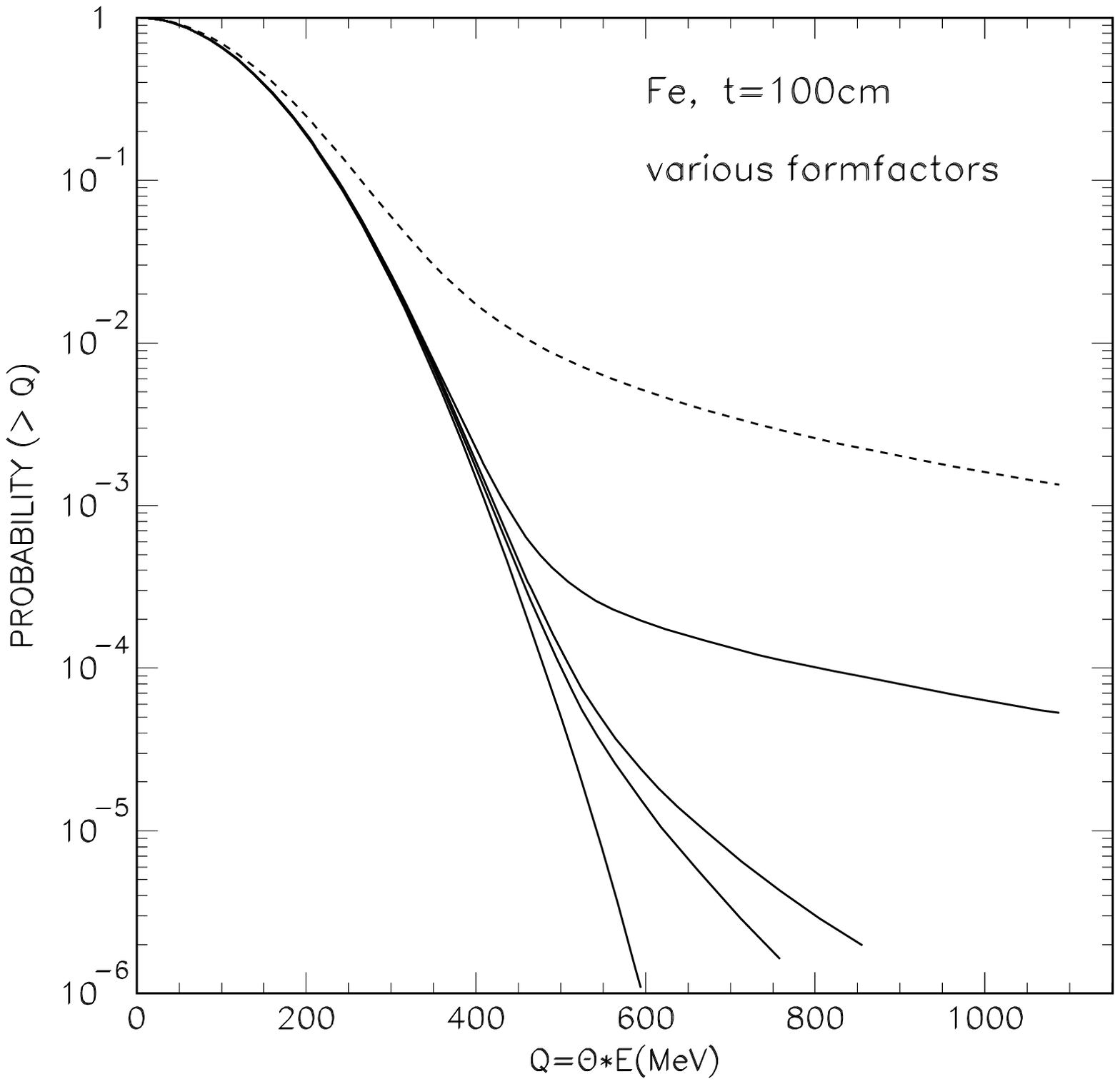,height=18cm,width=18cm}}

\vspace{2mm}
\noindent
\small
\end{center}
%%%%%%%%%%%%%%%%%%%%%%%%%%%%%%%%%
{\sf~~~~~~~~~~~~~~~~~~~~~~~~~~~~~~~~~~~~~~~~~~~~~~~~~~~~~~~~ Fig.~3}
%%%%%%%%%%%%%%%%%%%%%%%%%%%%%%%%%
\begin{center}
\mbox{\epsfig{file=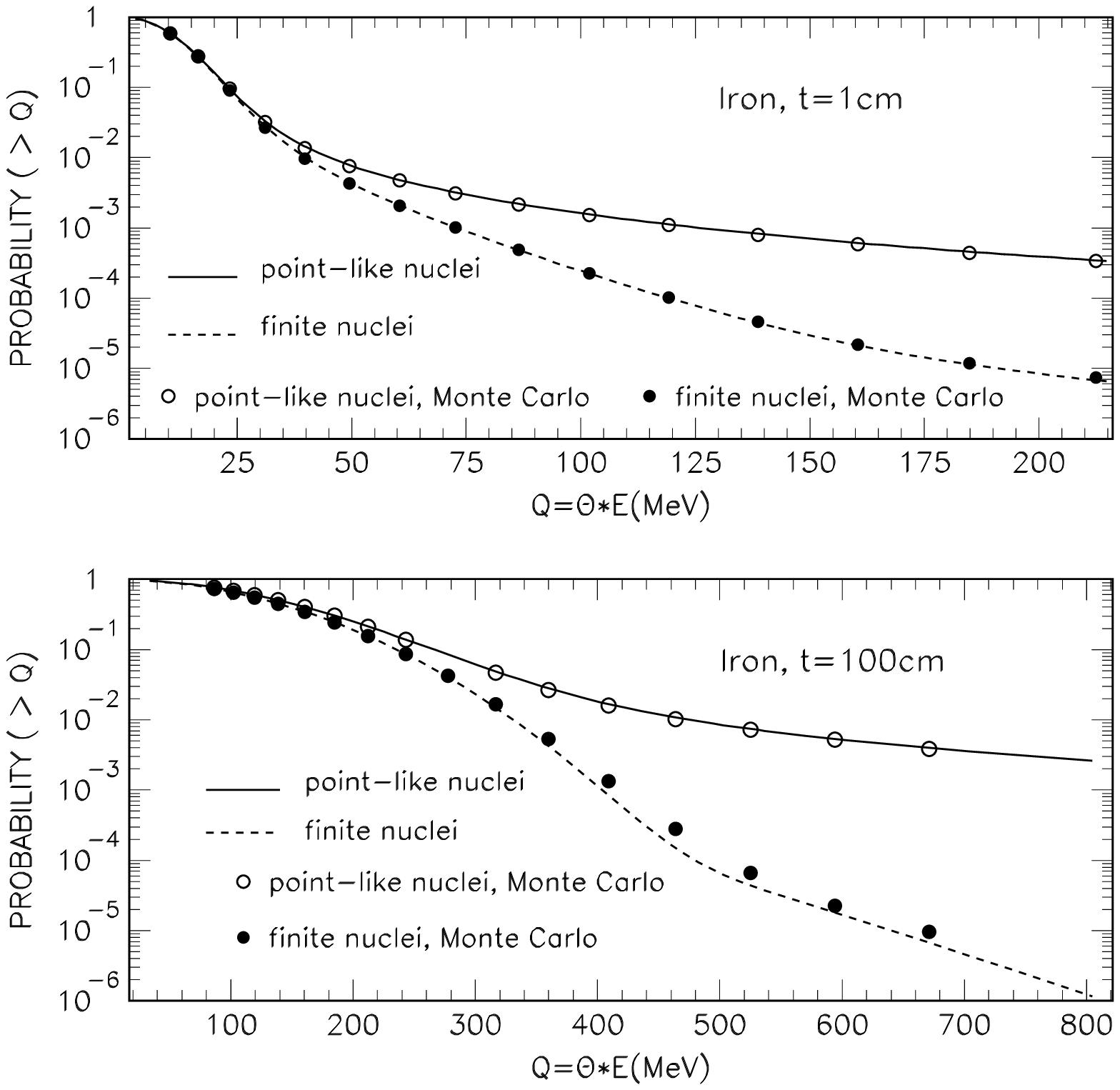,height=18cm,width=18cm}}

\vspace{2mm}
\noindent
\small
\end{center}
%%%%%%%%%%%%%%%%%%%%%%%%%%%%%%%%%
{\sf~~~~~~~~~~~~~~~~~~~~~~~~~~~~~~~~~~~~~~~~~~~~~~~~~~~~~~~~ Fig.~4}
%%%%%%%%%%%%%%%%%%%%%%%%%%%%%%%%%
\begin{center}
\mbox{\epsfig{file=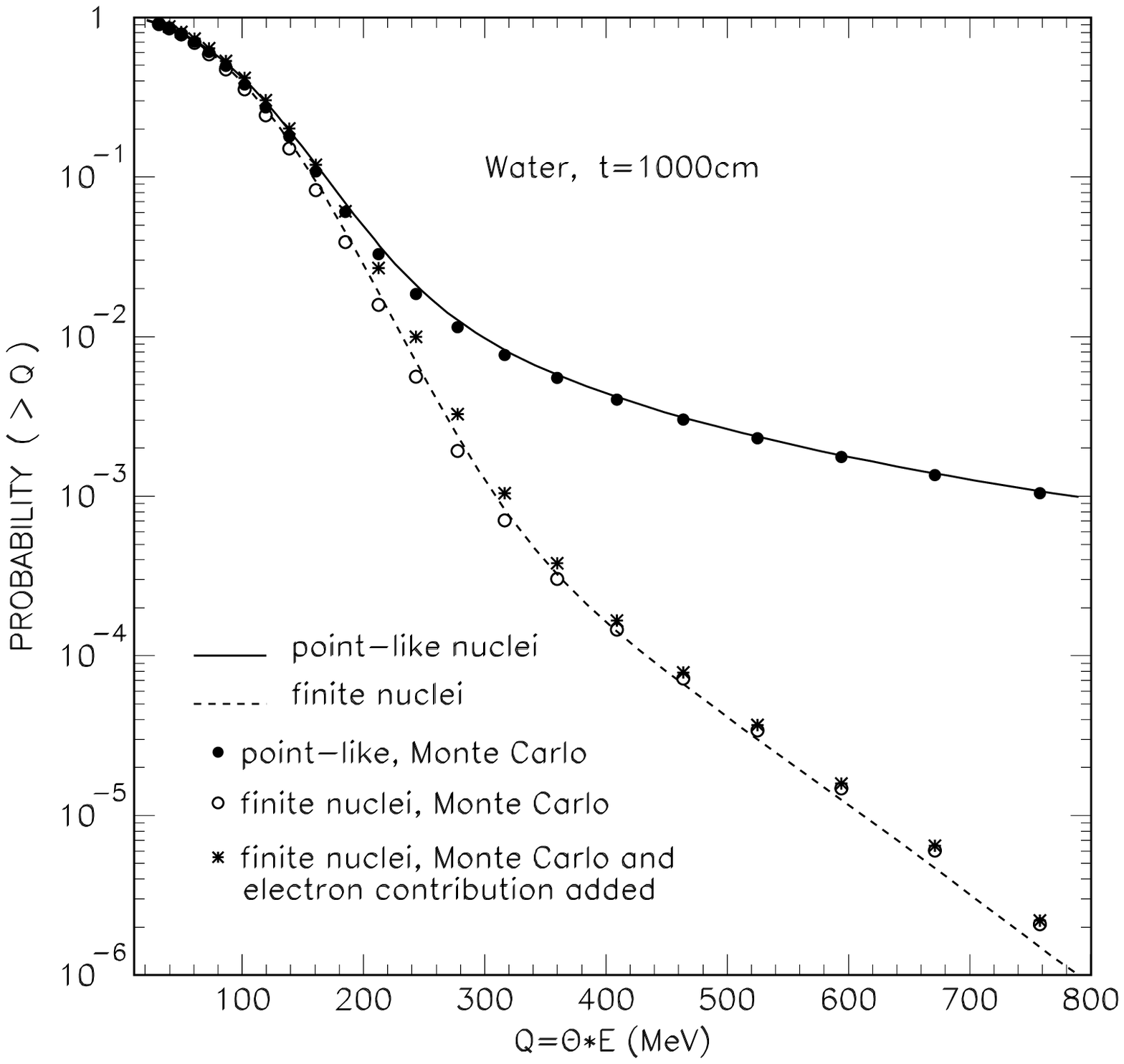,height=18cm,width=18cm}}

\vspace{2mm}
\noindent
\small
\end{center}
%%%%%%%%%%%%%%%%%%%%%%%%%%%%%%%%%
{\sf~~~~~~~~~~~~~~~~~~~~~~~~~~~~~~~~~~~~~~~~~~~~~~~~~~~~~~~~ Fig.~5}
%%%%%%%%%%%%%%%%%%%%%%%%%%%%%%%%%

\normalsize
\end{document}